\documentclass[%
 reprint,
 amsmath,amssymb,
 aps,
]{revtex4-2}

\usepackage{graphicx}
\usepackage{dcolumn}
\usepackage{bm}
\usepackage{array}
\usepackage{multirow}

\begin{document}

\preprint{APS/123-QED}

\title{Enhanced Extrapolation-Based Quantum Error Mitigation \\ Using Repetitive Structure in Quantum Algorithms}

\author{Boseon Kim}
 \email{boseon12@kisti.re.kr}
 \author{Wooyeong Song}%
 \email{wysong@kisti.re.kr}
 \author{Kwangil Bae}%
 \email{kibae@kisti.re.kr}
 \author{Wonhyuk Lee}%
 \email{livezone@kisti.re.kr}
 \author{IlKwon Sohn}%
 \email{d2estiny@kisti.re.kr}
\affiliation{%
Korea Institute of Science and Technology Information \\
Daejeon, 34141, Republic of korea
}%

\begin{abstract}
Quantum error mitigation is a crucial technique for suppressing errors especially in noisy intermediate-scale quantum devices, enabling more reliable quantum computation without the overhead of full error correction. Zero-Noise Extrapolation (ZNE), which we mainly consider in this work, is one of prominent quantum error mitigation methods. For algorithms with deep circuits - such as iterative quantum algorithms involving multiple oracle calls - ZNE's effectiveness is significantly degraded under high noise. Extrapolation based on such low-fidelity data often yields inaccurate estimates and requires substantial overhead. In this study, we propose a lightweight, extrapolation-based error mitigation framework tailored for structured quantum algorithms composed of repeating operational blocks. The proposed method characterizes the error of the repeated core operational block, rather than the full algorithm, using shallow circuits. Extrapolation is used to estimate the block fidelity, followed by a reconstruction of the mitigated success probability. We validate our method via simulations of the 6-qubit Grover’s algorithm on IBM’s Aer simulator, then further evaluating it on the real 127-qubit IBM Quantum system based on Eagle r3 under a physical noise environment. Our results, particularly those from Aer simulator, demonstrate that the core block's error follows a highly consistent exponential decay. This allows our technique to achieve robust error mitigation, overcoming the limitations of conventional ZNE which is often compromised by statistically unreliable data from near-random behavior under heavy noise. In low-noise conditions, our method approaches theoretical success probability, outperforms ZNE. In high-noise conditions, ZNE fails to mitigate errors due to overfitting of its extrapolation data, whereas our method achieves over a 20\% higher success probability.
\end{abstract}

\maketitle


\section{Introduction}
Computation based on the physical properties of qubits provides a theoretical advantage for solving certain problems more efficiently than classical methods~\cite{sh99, ar19}. Shor’s algorithm and Grover’s search algorithm are representative examples that demonstrate this potential. However, current quantum computers are Noisy Intermediate-Scale Quantum (NISQ) devices with tens to hundreds of qubits, and their computational accuracy is significantly limited by various sources of noise, including gate infidelity and measurement errors~\cite{pr18, bh22}. Quantum error mitigation (QEM) techniques are regarded as important approaches for addressing noise in the NISQ computing regimes~\cite{en21, ca23, he20, gi20, en18, vi99}. Various QEM techniques are actively being studied, including Zero-Noise Extrapolation (ZNE)~\cite{he20, gi20}, Probabilistic Error Cancellation (PEC)~\cite{en18}, and Dynamical Decoupling (DD)~\cite{vi99}. We focus on Zero-Noise Extrapolation (ZNE), a technique that estimates the ideal, noise-free expectation value by executing the same quantum circuit multiple times at different noise levels, amplified using unitary folding, and extrapolating the measured results. While ZNE is widely used in quantum computing, it may present the following issues in practical implementation. (i) As the noise scale increases, the circuit depth and the number of gates increase, which can lead to distorted expectation values or convergence toward random guessing due to accumulated quantum errors. (ii) Since each noise scale yields noisy measurement outcome with high statistical uncertainty, both multiple circuit executions and substantial data are required per scale. (iii) Expectation values at high noise scales with excessive noise insertion can lead to model overfitting during the extrapolation process, which may degrade the estimated ideal expectation value.

To address the above issues, various ZNE-related techniques have been proposed. The Layerwise Richardson Extrapolation (LRE) method decomposes a quantum circuit into gate-based layers and applies independent noise scaling to each layer~\cite{ru24}. By applying multivariate Richardson extrapolation, the method enhances the accuracy of ideal expectation value estimation and provides flexibility in handling noise variations across different layers. Random Identity Insertion Method (RIIM) reduces the gate overhead of Fixed Identity Insertion Method (FIIM)-based ZNE by inserting identity operations at random, and achieves the same extrapolation accuracy with fewer resources~\cite{he202}. The ZNE-based approaches described above have improved mitigation performance and efficiency by leveraging circuit structure, but methods that directly use the repeated block structure for error modeling remain relatively less explored.

This paper proposes a lightweight extrapolation-based error mitigation method that leverages the repeated structure of quantum algorithms with structural regularity, based on the progression of existing mitigation approaches. Structured quantum algorithms are characterized by the repeated use of core operational blocks, and leveraging this structural property enables efficient noise analysis and mitigation without extending the full circuit. In particular, errors occurring in the entire circuit can be modeled as the accumulation of block-level errors, allowing effective estimation of full circuit performance by quantifying the error characteristics of a single block. Consequently, we propose a lightweight error mitigation framework that integrates a estimation approach based on block-level repetition into existing ZNE techniques, aiming to reduce resource overhead and enhance success probability. This paper is organized as follows. Section II introduces an overview of structured quantum algorithms and Grover's search algorithm, which serves as a representative example and is the primary algorithm used in this study. A brief overview of the ZNE is also provided. Section III describes the proposed method. Section IV presents a comparative analysis of the proposed method and conventional ZNE using IBM’s Aersimulator and the IBM Eagle r3 quantum system.

\section{Preliminaries}
This section introduces the concept of structured quantum algorithms and explains Grover’s search algorithm as a representative example used in this study. It then describes conventional ZNE techniques for comparison with the proposed lightweight error mitigation method.

\subsection{\label{sec:level2}Structured Quantum Algorithms}
In this study, we define quantum algorithms with repetitive structures as structured quantum algorithms. In such algorithms, an operational block with a repeated structure is defined as the core operational block. structured quantum algorithms contain a core operational block that is repeated $r$ times, where $r$ is arbitrary.

A representative example of a structured quantum algorithm is Grover’s search algorithm, which consists of Hadamard ($H$), oracle ($O$), diffusion ($D$), and measurement ($M$) operations~\cite{gr96}. The algorithm amplifies the probability of the target state by repeating a sequence of oracle and diffusion operations $r$ times. Grover’s search algorithm applies the $DO$ block repeatedly for the optimal number of iterations $r_\mathrm{opt}$, and for an $n$-qubit system, it can be expressed as \( (DO)^{r_{\mathrm{opt}}} H^{\otimes n} \). As shown in Fig1, this study defines the operational block consisting of $O$ and $D$ operators in Grover’s search algorithm as the core operational block.

\noindent
\begin{figure}[t]
    \includegraphics[width=0.45\textwidth]{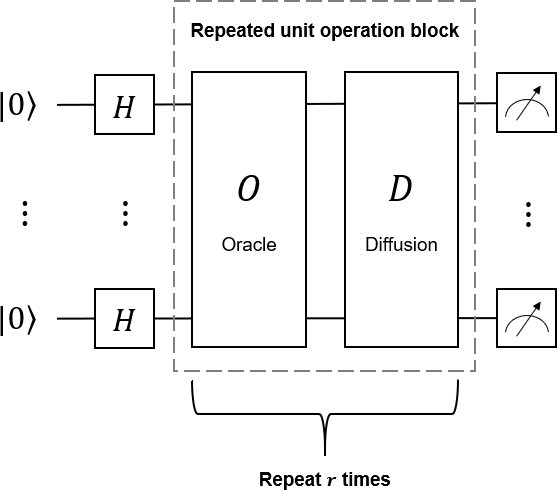}
    \caption{A representative example of a structured quantum algorithm is Grover’s search: the operational block consisting of the oracle $O$ and diffusion $D$ operators is defined as the core operational block.}
    \label{fig:my_figure}
\end{figure}

This section introduced the structured quantum algorithm as defined in this study and defined the core operational block through a representative example. It then describes the ZNE technique selected as a baseline for comparison with the proposed method.

\subsection{\label{sec:level2}Zero-Noise Extrapolation}
ZNE is the one of the leading error mitigation techniques that does not directly eliminate quantum errors but instead estimates the ideal, noise-free expectation value through extrapolation. The overall procedure of ZNE can be divided into the following three steps.

First, noise scaling involves repeatedly executing the same quantum circuit at different noise scales. A representative method for noise scaling is unitary folding, which artificially increases the noise by expanding quantum circuits or gates as follows~\cite{pr22}.
\begin{equation}
U \rightarrow U \cdot U^\dagger \cdot U.
\end{equation}

Theoretically, $UU^\dagger$ yields the identity operator $I$ according to the unitary condition, but in practice, noise accumulates as the number of gates increases on actual hardware. The noise scale, denoted as $\lambda$, represents both the number of circuit repetitions and the degree of noise amplification. Noise scaling is expressed as follows, where $n$ denotes the number of folded $UU^\dagger$ pairs.
\begin{equation}
U^{\lambda} = U (U^{\dagger} U)^n,
\end{equation}
where $\lambda =2n+1, \quad n \in \mathbb{Z}_{\geq 0}$.

Second, for each noise scale $\lambda$, perform $N_{\text{shots}}$ repeated measurements to obtain the number of target state outcomes $N_{\lambda,\mathrm{target}}$. At noise scale $\lambda$, the target expectation value $E(\lambda)$ is given by
\begin{equation}
E(\lambda) = \frac{N_{\lambda,\text{target}}}{N_{\text{shots}}}.
\end{equation}

Third, the expectation values $E(\lambda)$ measured at each noise scale $\lambda$ are extrapolated to $\lambda=0$ to estimate the noise-free expectation value $E(0)$. Common extrapolation methods include polynomial, Richardson, log–linear, and exponential extrapolation~\cite{br13, si03}.

Among the various extrapolation methods, this paper uses exponential, Richardson, and log–linear extrapolation to estimate the noise-free expectation value $E(\lambda)$. These methods are based on the assumption that noise accumulates exponentially in the circuit. Under the assumption that the measured value $E(\lambda)$ decays exponentially with $\lambda$, the ideal expectation value $E(0)$ is estimated. Exponential extrapolation assumes that
\begin{equation}
E(\lambda) = a + b p^{\lambda}.
\end{equation}
where $p$ denotes the decay rate due to noise. Log–linear extrapolation applies a logarithmic transformation to exponential extrapolation and is expressed by the following equation.
\begin{equation}
\log(E(\lambda) - a) = \log b + \lambda \log p.
\end{equation}

These techniques enable more accurate estimation compared to conventional linear extrapolation methods, as they effectively capture the characteristic behavior of quantum circuits where the expectation value rapidly decreases under high noise levels.  

Based on the structured algorithms and ZNE techniques introduced in this section, the following section presents the proposed error mitigation method.

\section{Method}
This chapter presents the proposed extrapolation-based error mitigation method for structured quantum algorithms. The methodology proceeds as follows. First, we extract the core operational block repeatedly used in the structured quantum algorithm and characterize its error to compute the corrected success probability of the full circuit. This is followed by a detailed, step by step explanation.

Let $U_{\text{total}}$ include a core operational block $U_{\text{iter}}$ repeated $r$ times. when measuring $U_{\text{total}}$ for $N_{\text{shots}}$ shots, the raw success probability $P_{r,\mathrm{raw}}$ of observing the target state is 

\begin{equation}
P_{r,\mathrm{raw}} = \frac{N_{\mathrm{raw},\mathrm{target}}}{N_{\mathrm{shots}}},
\end{equation}
where $N_{\mathrm{raw},\mathrm{target}}$ is the number of times the target state was observed.

To quantify the error characteristics of $U_{\text{iter}}$, we apply $U_{\text{iter}}$ followed by its inverse $U^{\dagger}_{\text{iter}}$ in sequence to derive the identity operation $I$. One such pair is denoted as
\begin{equation}
U_I = U^{\dagger}_{\mathrm{iter}} U_{\mathrm{iter}} = I.
\end{equation}

When the two-block sequence $U_I$ is folded $k$ times, the resulting operator $(U_I)^k$ implements the identity operator $I$.
\begin{equation}
(U_I)^k = \left( U^{\dagger}_{\mathrm{iter}} \, U_{\mathrm{iter}} \right)^k = I.
\end{equation}

$(U_I)^k$ is executed over $N_{\text{shots}}$ repetitions, and the number of times $N_{\mathrm{init}}$ that the initial state $\left| \psi_0 \right\rangle$ is observed is counted. Using $N_{\mathrm{init}}$, the fidelity of the two-block sequence $U_I$ for each $k$ is calculated. The following equation defines $F_I(2k)$, the probability of returning to the initial state:
\begin{equation}
F_I(2k) = \frac{N_{\mathrm{init}}(2k)}{N_{\mathrm{shots}}},
\end{equation}
where $2k = 2,\, 4,\, \cdots,\, 2\left\lfloor \frac{r}{2} \right\rfloor$.

We estimate the fidelity of the single core operational block $U_{\text{iter}}$ from the fidelity of the two-block sequence $U_I$ at each $k$. To estimate the fidelity of $U_\mathrm{I} = U^{\dagger}_{\mathrm{iter}} U_{\mathrm{iter}}$, we use the exponential decay model commonly employed in ZNE, in which $F_I(2k)$ denotes the fidelity when two block sequence $U_\mathrm{I}$ is applied $k$ times. This model is based on the assumption that noise accumulates with each repetition $k$, causing the fidelity to decay exponentially. The model is expressed as follows:
\begin{equation}
F_I(2k) = c f^{2k}.
\end{equation}

The exponential decay model assumes that the fidelity decays as a function of the number of block repetitions. $F(k)$ denotes the fidelity when the core operational block $U_\mathrm{iter}$ is applied $k$ times. It is expressed as
\begin{equation}
F(k) = c\,f^{k}, \quad F_I(2k) = c\,f^{2k}.
\end{equation}

$c$ is the initial state fidelity—including state preparation and measurement loss—and $f\in[0,1]$ is the ideal fidelity retained by a single core operational block $U_{\mathrm{iter}}$. $f$ denotes the preserved fidelity each time the block is applied, and the fidelity of full circuit decays exponentially by a factor of $f^k$ after k repetitions. $F_I(2k)$ denotes the fidelity of the full circuit after $k$ repetitions of $U_I$. The values $F_I(2k)$ are obtained for each $k$ and then the exponential decay model to determine $f$ and $c$. Depending on the value of $c$, different formulas are used to calculate the fidelity of the single block $U_{\text{iter}}$. If $c$ converges to 1, $f= \sqrt[2k]{F_I(2k)}$.

For $c \neq 1$ and $k = 1$ and $2$, $f$ and $c$ are determined using the following equation:
\begin{equation}
\frac{F_I(4)}{F_I(2)} = \frac{c f^4}{c f^2} = f^2.
\end{equation}

For at least three $k$ values, $f$ and $c$ are determined by log–linear extrapolation. For $k \geq 3$, discard any point where $F_I(2k)$ is indistinguishable from the uninformative baseline set by effective noise and readout errors, to avoid overfitting to baseline‑dominated measurements. For example, using $k=1,2,3$, $f$ and $c$ are computed as follows.
\begin{align}
\ln F(k) = \ln c + k \ln f,\\
\ln f = \frac{(\ln F(6) - \ln F(2))}{4},\\
\ln c = \ln F(2) - 2 \ln f.
\end{align}

We correct the ideal success probability of the full circuit using the single-block fidelity $f$. Instead of the experimentally measured success probability $P_{r,\mathrm{raw}}$, we estimate the noise-free success probability $P_{\mathrm{ideal}}$. When $U_{\mathrm{total}}$ is measured for $N_{\mathrm{shots}}$ shots, the probability of obtaining the target state is $P_{r,\mathrm{raw}}$. 
If $f$ is the fidelity of the single block $U_{\mathrm{iter}}$, then the raw success probability follows,
\begin{equation}
P_{r,\mathrm{raw}} \approx P_{\mathrm{ideal}} \times c f^r.
\end{equation}

i.e.\ the raw success probability decays exponentially with $r$ due to repeated fidelity loss of factor $f$. Based on the exponential decay model, the error-mitigated success probability $P_{\mathrm{mit}}$ is given by:
\begin{equation}
P_{\mathrm{mit}} \approx \frac{P_{r,\mathrm{raw}}}{c f^r}.
\end{equation}

\section{Performance Evaluation}
This section presents three evaluations. First, we compare the error-mitigated success probabilities of the proposed method and standard ZNE on IBM’s Aersimulator with added arbitrary noise. Second, after validating the method on the Aersimulator, we execute the method on IBM's Eagle r3 quantum processor to validate performance improvements on real quantum system. Third, to assume a future quantum system environment with improved error rates, the proposed method is executed on the Aersimulator under further reduced-noise conditions.

\subsection{\label{sec:level2}Performance Analysis of the 6‑Qubit Grover Algorithm Using the Aersimulator}
A representative example of a structured quantum algorithm is Grover’s search. We define the identity block for the $OD$ operation in $U_{\mathrm{total}}$ as $U_I = (OD)^\dagger OD = I$. For the 6‑qubit Grover search we use, the optimal iteration count is $r_{\mathrm{opt}} = 6$, requiring six iterations of the OD block to amplify the probability of the target state $\lvert111111\rangle$. Theoretically, for a 6‑qubit search over 64 basis states, the success probability after six iterations is approximately 99.7\%.

Each circuit is executed with $N_{\mathrm{shots}}=4000$ shots per run, and this procedure is repeated independently 10 times to ensure statistical confidence. In the IBM Aersimulator, we apply a depolarizing noise model with error probability $p=10^{-4}$ on every single‑qubit gate and $p=10^{-3}$ on every two‑qubit gate.
Across 10 independent runs of 4000 shots each, the target state $\lvert111111\rangle$ was observed an average of 1679 times, corresponding to an unmitigated success probability of approximately 41.975\%.
$P^{(i)}_{6,\mathrm{raw}}$ is the raw success probability observed in the $i$th run of the 6‑qubit circuit, and $\overline{P}_{6,\mathrm{raw}}$ is the average over 10 runs.

For $k = 1, 2, 3$, $(U_I)^k$ is measured over 4000 shots, and the fidelity of the two-block sequence $U_I$ at each $k$ is evaluated based on the number of times the initial state $\lvert000000\rangle$ is observed.
\begin{table}[h]
\centering
\renewcommand{\arraystretch}{1.2}
  \begin{tabular}{|
  >{\hspace{6pt}}c<{\hspace{6pt}} |
  >{\hspace{6pt}}c<{\hspace{6pt}} |
}
    \hline
    \textbf{The number of} & \textbf{Fidelity} \\[-3pt]
    \textbf{repetitions ($k$)} & \textbf{($\overline{F_I(2k)}$)} \\
    \hline
    \rule{0pt}{2.5ex}1 &  $\overline{F}_I(2) = 0.755$ \\
    \hline
    \rule{0pt}{2.5ex}2 &  $\overline{F}_I(4) = 0.57$ \\
    \hline
    \rule{0pt}{2.5ex}3 &  $\overline{F}_I(6) = 0.436$ \\
    \hline
  \end{tabular}
\caption{Average fidelity per $(OD)_I$ block depending on the number of repetitions}
\label{tab:fidelity_by_repetition}
\end{table}

The fidelity of the single‑block OD is estimated by log–linear extrapolation of the exponential decay model. In these experiments, depolarizing errors are applied to each gate in the Aersimulator; SPAM errors are excluded, and the initial state fidelity $c$ is therefore assumed to be 1.

Over ten independent runs, the fidelities $F_I(2)$, $F_I(4)$, and $F_I(6)$ were computed and averaged to yield $\bar f = 0.8693$. Using $\bar f$, the error-mitigated success probability $\overline{P}_{\text{mit}}$ is $0.9686$.

Given at least three $F(k)$ values, $f$ is determined via log–linear extrapolation. But the same data can also be analyzed by taking the appropriate root of each $F_I(2k)$. The block fidelity estimates by root extraction are:
\begin{table}[h]
\centering
\renewcommand{\arraystretch}{1.4}
\begin{tabular}{|
  >{\hspace{6pt}}c<{\hspace{6pt}} |
  >{\hspace{6pt}}c<{\hspace{6pt}} |
  >{\hspace{6pt}}c<{\hspace{6pt}} |
}
\hline
\rule{0pt}{2.8ex}\textbf{$k$} & \textbf{$\overline{F_I(2k)}$} & \textbf{$f$} \\
\hline
\rule{0pt}{2.5ex}1 & 0.7549 & $\sqrt{0.7549} \approx 0.8689$ \\
\hline
\rule{0pt}{2.5ex}2 & 0.57015 &$\sqrt[4]{0.57015} \approx 0.8690$ \\
\hline
\rule{0pt}{2.5ex}3 & 0.43658 &$\sqrt[6]{0.43658} \approx 0.871$ \\
\hline
\end{tabular}
\caption{Average fidelity $f$ based on exponential decay analysis}
\end{table}
That is, the single‑block $OD$ fidelity $f$ converges to 0.8696, and $F(k)$ decays exponentially as $F(k)\propto f^k$ with increasing $k$. Although $f$ is in principle determined via log–linear extrapolation, the measured probability of the target state $\lvert000000\rangle$ exhibited a clear exponential pattern for each $k$. Therefore, simple root‑extraction analysis yields an estimate of $f$ converging to $0.8696 \pm 0.002$. By characterizing gate errors at the block level and applying this model to Grover’s search on the Aersimulator, we estimate $f$ using simple root‑extraction analysis. This demonstrates that the exponential decay model assumed by the proposed method is well matched to both Grover’s algorithm and the Aersimulator. 

\begin{table*}[t]
\centering
\renewcommand{\arraystretch}{1.5} 
\begin{tabular}{|c|c|c|c|c|c|c|c|c|c|c|c|}
\hline
\multirow{2}{*}{\textbf{qubit}} &
\multirow{2}{*}{\textbf{theoretical}} &
\multirow{2}{*}{\textbf{unmitigated}} &
\multicolumn{4}{c|}{\textbf{ZNE}} &
\multicolumn{5}{c|}{\textbf{Proposed}} \\
\cline{4-12}
& & & \textbf{E(1)} & \textbf{E(3)} & \textbf{E(5)} & $\bm{P_{zne}}$ &
\textbf{F(2)} & \textbf{F(4)} & \textbf{f} & \textbf{c} & $\bm{P_{mit}}$ \\
\hline
3 & 0.945 & 0.675 & 0.676 & 0.366 & 0.18 & 0.877 & 0.642 & 0.428 & 0.816 & 0.907 & 0.970 \\
\hline
4 & 0.961 & 0.109 & 0.098 & 0.061 & (0.063) & 0.117 & 0.201 & 0.082 & 0.662 & 0.467 & 0.907 \\
\hline
\end{tabular}
\caption{Measurement success probabilities and related parameters for IBM-based 3- and 4-qubit Grover search algorithms.}
\label{tab:zne_proposed_comparison}
\end{table*}

To compare with the proposed method, we apply standard ZNE to Grover's search algorithm. For each noise-scale factor $\lambda$, the circuit is folded $\lambda$ times, and the target expectation value $E(\lambda$ is measured. The result for $E(1)= 0.41868$, $E(3)= 0.09045$, $E(5)= 0.0312$.
By extrapolating the expectation values over different noise scales, we obtain an error‑mitigated success probability of approximately 68.37\%.
In this study, we denote the ZNE-mitigated success probability as $P_\text{zne}$.

\noindent
\begin{figure}[t]
    \includegraphics[width=0.45\textwidth]{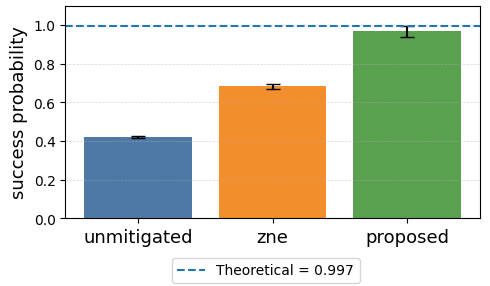}
    \caption{Success probability of the 6-qubit Grover search algorithm on the Aersimulator for different methods: The theoretical value of 99.7\% is reduced to 41.975\% without mitigation. Standard ZNE restores this to 68.37\%, but at noise scale $\lambda=5$ the extrapolated value approaches the random-guessing probability, risking overestimation. In contract, the proposed method recovers the error-mitigated success probability to 96.86\%, closely matching the theoretical value.}
    \label{fig:my_figure}
\end{figure}

Overall, the theoretical success probability of the 6‑qubit Grover algorithm is 99.7\% after $r_\mathrm{opt}$ iterations, but noise reduces its unmitigated success probability to 41.98\%, which corresponds to a 57.72\% decrease from the theoretical value. Applying standard ZNE raises the success probability to 68.37\%, partially recovering this loss; however, at noise scale $\lambda = 5$, the expectation value drops to 3.1\%, approaching the level of random guessing $(1/64 \approx 1.6\%)$. Including low‑confidence data in the extrapolation poses a risk of significantly overestimating the error. In contrast, our block‑level error characterization method increases the success probability to 96.86\%, approaches the ideal more closely, and provides more efficient and robust error mitigation than standard ZNE.

\subsection{\label{sec:level2}Performance Analysis of the 3,4‑Qubit Grover Algorithm Using the IBM Eagle r3}

In the previous chapter, we validated the proposed method on the Aersimulator. In this chapter, we test the performance of the proposed method under physical noise on IBM’s Eagle r3 quantum processor. We test the 3‑qubit and 4‑qubit Grover search algorithms, with optimal iteration counts $r_\mathrm{opt}=2$, $3$, respectively. Each circuit is executed with $N_\text{shots}=4000$ per run, and this procedure is repeated independently 10 times to ensure statistical confidence. 

Table II presents the average results over ten independent runs. For the 3‑qubit Grover search targeting \(\lvert111\rangle\) among eight computational basis states with \(r_{\mathrm{opt}}=2\), the theoretical success probability is 94.5\%. The unmitigated success probability is 67.5\%, indicating degradation from accumulated gate and SPAM errors. Applying standard ZNE and the proposed method restores the error-mitigated success probability closer to the theoretical value. When $P_\text{mit}$ and $P\text{zne}$ exceed 1, they are clipped to 1 to enforce the physical bound.

\noindent
\begin{figure}[t]
    \includegraphics[width=0.45\textwidth]{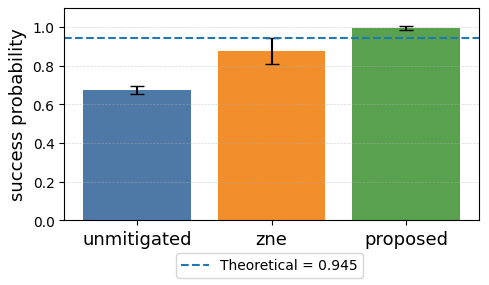}
    \caption{Success probability of the 3-qubit Grover search algorithm on the IBM hardware for different methods: The theoretical value of 94.5\% is reduced to 67.5\% without mitigation, but is restored near the theoretical value after applying ZNE and the proposed method.}
    \label{fig:my_figure}
\end{figure}

Table III presents the average results over ten independent runs. For the 4‑qubit Grover search targeting \(\lvert1111\rangle\) among sixteen computational basis states with \(r_{\mathrm{opt}}=3\), the theoretical success probability is 96.1\%. The unmitigated success probability is 10.9\%, indicating degradation from accumulated gate and SPAM errors. Even with ZNE, the error-mitigated success probability reaches only 12.7\%. Because $\lambda=3$ and $5$ match the random-guessing probability, they cannot be used reliably in the extrapolation. In contrast, our method raises the mitigated success probability $P_\text{mit}$ from the unmitigated success probability of 10.9\% to 90.7\% despite a single‑block fidelity $f \approx 0.66$ and initial state fidelity $c \approx 0.49$. This result approaches the theoretical success probability and demonstrates that block‑level error characterization remains effective as circuit depth increases.

\noindent
\begin{figure}[t]
    \includegraphics[width=0.45\textwidth]{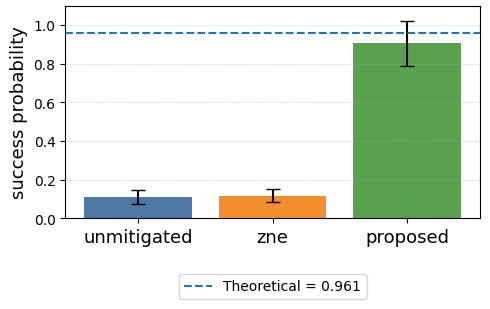}
    \caption{Success probability of the 4-qubit Grover search algorithm on the IBM hardware for different methods: The theoretical value of 96.1\% is reduced to 11.7\% without mitigation; standard ZNE yields only 11.7\% due to overestimation from low-confidence data, whereas the proposed method restores performance close to the theoretical value.}
    \label{fig:my_figure}
\end{figure}

\begin{table*}[t]
\centering
\renewcommand{\arraystretch}{1.5} 
\begin{tabular}{|c|c|c|c|c|c|c|c|c|c|c|c|c|}
\hline
\multirow{2}{*}{\textbf{qubit}} & \multirow{2}{*}{\textbf{error rate}} & \multirow{2}{*}{\textbf{theoretical}} & \multirow{2}{*}{\textbf{unmitigated}} &
\multicolumn{4}{c|}{\textbf{ZNE}} & \multicolumn{5}{c|}{\textbf{Proposed}} \\
\cline{5-13}
& & & & \textbf{E(1)} & \textbf{E(3)} & \textbf{E(5)} & $\bm{P_{zne}}$
& \textbf{F(2)} & \textbf{F(4)} & \textbf{F(6)} & \textbf{f} & $\bm{P_{mit}}$ \\
\hline
\multirow{1}{*}{5} & 0.005 & 0.999 & 0.295 & 0.292 & 0.053 & (0.032) & 0.41 & 0.55 & 0.31 & -- & 0.751 & 0.93 \\
\hline
\multirow{3}{*}{6} & 0.005 & -- & 0.032 & 0.031 & 0.016 & (0.015) & 0.039 & 0.263 & 0.086 & (0.039) & 0.572 & 0.299 \\
& 0.001 & 0.9635 & 0.42 & 0.419 & 0.09 & 0.03 & 0.684 & 0.755 & 0.57 & 0.437 & 0.87 & 0.969 \\
& 0.0005 & -- & 0.645 & 0.643 & 0.278 & 0.129 & 0.908 & 0.868 & 0.758 & 0.659 & 0.933 & 0.979 \\
\hline
\end{tabular}
\caption{Performance analysis of 5- and 6-qubit Grover algorithm under error rates based on the Aersimulator.}
\label{tab:aersim_error_rate}
\end{table*}

\subsection{\label{sec:level2}Performance Analysis of the 5,6‑Qubit Grover Algorithm under gate error rates Using the Aersimulator}

In the previous chapter, we evaluated the proposed method on existing quantum processer, in this chapter we simulate enhanced quantum processor performance by reducing the gate error rate in the Aersimulator and evaluate the proposed method under these conditions. The depolarizing error model reproduces error patterns similar to those observed in real quantum processor. We set the two-qubit gate error rates to 0.005, 0.001, and 0.0005, with single-qubit gate error at 0.0005, 0.0001, and 0.00005, respectively. Under these conditions, we apply the 5 and 6-qubit Grover's algorithms to evaluate the proposed method’s robustness to noise variations and the consistency of its mitigation effect. Hereafter, we refer to the two‑qubit gate error rate simply as the error rate.

 First, we evaluate the performance of the 5‑qubit Grover search algorithm under the error rate of 0.005. The theoretical success probability is 99.9\%, but the unmitigated success probability is only 29.5\%. Applying standard ZNE increases this to 41\%, still far from the theoretical value. At noise scale $\lambda=5$, $E(5)$ falls to the random guessing probability $(1/32 \approx 0.0312)$ risking overestimation in the extrapolation. Consequently, data points which success probability falls near random guessing is discarded before extrapolation. Using our method, the error-mitigated success probability reaches 93\%, closely approaching the theoretical success probability.

\noindent
\begin{figure}[t]
    \includegraphics[width=0.45\textwidth]{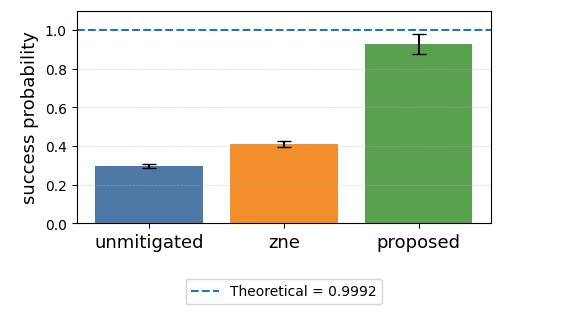}
    \caption{Success probability of the 5-qubit Grover search algorithm on the Aersimulator for different methods and error rate: standard ZNE corrects the rate to 41\%, which remains far from the theoretical value; in contrast, the proposed method achieves the error-mitigated success probability of 93\%, closely matching the theoretical value.}
    \label{fig:my_figure}
\end{figure}

Second, we evaluate the performance of the 6‑qubit Grover search algorithm under the error rate of 0.005, 0.001, 0.0005. The results are shown in Fig.X and Table III. At an error rate of 0.005, $E(5)$ and $F_I(6)$ approximate the random-guessing probability $(1/64 \approx 0.0156)$ and are excluded from the extrapolation. As the error rate increases, the unmitigated success probability correspondingly decreased. At lower rates of $0.001$ and $0.00005$, standard ZNE increases the error-mitigated success probability close to the theoretical value, but under the high noise of 0.005 it remains near the unmitigated value. In contrast, our method achieves near‑theoretical success probability at 0.001 and 0.0005 and still provides substantial error-mitigation at 0.005, demonstrating strong robustness. Therefore, although standard ZNE can reliably mitigate errors at low noise, it fails under high noise. However, our block‑level characterization method, remains robust under fluctuating noise environments and is expected to perform effectively on future quantum processors with high error rates.

\noindent
\begin{figure}[t]
    \includegraphics[width=0.45\textwidth]{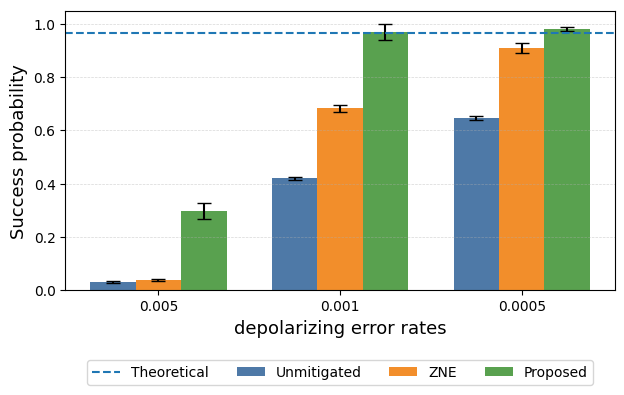}
    \caption{Success probability of the 6-qubit Grover search algorithm on the Aersimulator for different methods and error rate: Both standard ZNE and the proposed method approach the theoretical success probability at low noise, but only the proposed method provides significant recovery at the high noise.}
    \label{fig:my_figure}
\end{figure}

\section{Conclusion}
This paper proposes a lightweight, extrapolation‑based error mitigation technique for structured quantum algorithms. Repeated core operational block is assembled into the identity operation block, and block fidelity is estimated by fitting an exponential decay model. The estimated fidelity is then used to correct error-mitigated success probability, reconstructing the theoretical success probability. In low noise conditions, our method approaches the theoretical success probability and outperforms ZNE in error mitigation. In high noise conditions, the data used for ZNE extrapolation approaches the random-guessing value, leading overfitting during extrapolation. By contrast, our method achieves more than a 20\% higher success probability than ZNE under the same conditions.

\begin{acknowledgments}
We are grateful to Dr. J. Bang and Dr. K. Baek for insightful discussions and for their kind assistance related to the IBM Quantum System One at Yonsei.
This work was supported by Quantum Computing based on Quantum Advantage challenge research through the National Research Foundation of Korea (NRF) funded by the Korean government (MSIT) (RS-2023-00256221).
\end{acknowledgments}

\providecommand{\noopsort}[1]{}\providecommand{\singleletter}[1]{#1}%

\end{document}